\documentclass[usenatbib,twocolumn, trackchanges]{aastex63}

\usepackage{graphicx}
\usepackage{hyperref}
\usepackage{natbib}
\usepackage{xcolor,colortbl}
\usepackage{enumitem}
\usepackage{amsmath}
\usepackage{lineno}
\usepackage{color,soul}

\newcommand{\mjup}{$M_{\rm Jup}$}

\shorttitle{Brown Dwarf Microlensing}

\begin{document}

\title{A Search for Predicted Astrometric Microlensing Events by Nearby Brown Dwarfs}

\author[0000-0003-4422-2003]{Judah Luberto}
\email{judah@astro.ucla.edu}
\affiliation{Department of Astronomy and Astrophysics, University of California, Santa Cruz, CA 93105, USA}
\affiliation{Department of Physics and Astronomy, University of California, Los Angeles, CA 90024, USA}

\author[0000-0002-0618-5128]{Emily C. Martin}
\affiliation{Department of Astronomy and Astrophysics, University of California, Santa Cruz, CA 93105, USA}

\author[0000-0002-1052-6749]{Peter McGill}
\affiliation{Department of Astronomy and Astrophysics, University of California, Santa Cruz, CA 93105, USA}
 
\author{Alexie Leauthaud}
\affiliation{Department of Astronomy and Astrophysics, University of California, Santa Cruz, CA 93105, USA}
 
\author[0000-0001-6098-3924]{Andrew J. Skemer}
\affiliation{Department of Astronomy and Astrophysics, University of California, Santa Cruz, CA 93105, USA}
 
\author[0000-0001-9611-0009]{Jessica R. Lu}
\affiliation{Department of Astronomy, University of California, Berkeley, CA 94720, USA}

\begin{abstract}
\noindent Gravitational microlensing has the potential to provide direct gravitational masses of single, free-floating brown dwarfs, independent of evolutionary and atmospheric models. The proper motions and parallaxes of nearby brown dwarfs can be used to predict close future alignments with distant background stars that cause a microlensing event. Targeted astrometric follow up of the predicted microlensing events permits the brown dwarf’s mass to be measured. Predicted microlensing events are typically found via searching for a peak threshold signal using an estimate of the lens mass. We develop a novel method that finds predicted events that instead will lead to a target lens mass precision. The main advantage of our method is that it does not require a lens mass estimate. We use this method to search for predicted astrometric microlensing events occurring between 2014 - 2032 using a catalog of 1225 low mass star and brown dwarf lenses in the Solar Neighborhood of spectral type M6 or later and a background source catalog from DECaLS Data Release 9. The background source catalog extends to $g = $ 23.95, providing a more dense catalog compared to \textit{Gaia}. Our search did not reveal any upcoming microlensing events. We estimate the rate of astrometric microlensing event for brown dwarfs in the Legacy Survey and find it to be low $\sim10^{-5}$yr$^{-1}$. We recommend carrying out targeted searches for brown dwarfs in front of the Galactic Bulge and Plane to find astrometric microlensing events that will allow the masses of single, free-floating brown dwarfs to be measured.
\end{abstract}


\section{Introduction}
A brown dwarf's mass is its most fundamental physical attribute. Mass sets the evolutionary path of an individual brown dwarf as it contracts and cools over time \citep{burrows1997}. Surface gravity ($g \propto M/R^2$) plays a large role in determining the bulk atmospheric properties and atmospheric evolution \citep{marleyrobinson2015} of brown dwarfs. On a population level, brown dwarf masses span the gap between stars and gas giant planets so their masses inform studies of both the low mass end of the star formation process and the high mass end of giant planet formation \citep[e.g.,][]{chabrierbaraffe2000}.

The continual evolution of a brown dwarf's effective temperature means that no main sequence exists to easily determine a brown dwarf's mass. For brown dwarfs in bound multiple systems, dynamical masses can be measured by directly imaging the orbital dynamics \citep[e.g.,][]{konopacky2010, dupuyliu2012} or indirectly determining the dynamical masses, via a combination of astrometric and RV measurements \citep[e.g.,][]{brandt2019}. However, the masses of free-floating brown dwarfs are typically inferred by comparisons to evolutionary models based on measured parallax and bolometric luminosity, which can provide mass estimates with typical precision of 10--20 \mjup \citep[e.g.,][]{filippazzo2015, leggett2017}.

Gravitational microlensing events can offer a way to directly measure the mass of single, free floating brown dwarfs  \citep{Paczynski1996,Evans2014, cushing}. All that is required for a mircolensing event is the chance alignment between a foreground lens (a brown dwarf in our case) and a more distant background star. In principle microlensing events can have both photometric \citep{Refsdal1964,Paczynski1986} and astrometric signals \citep{walker,Hog1995,Miyamoto1995}.  Typically, microlensing events are found photometrically by large-scale monitoring surveys of the Galactic bulge and plane \citep[e.g.,][]{Udalski2015, Kim2016, Husseiniova2021}. For the vast majority of photometric microlensing events currently detected, the mass of the lens is unconstrained. Only those events with higher-order effects (e.g., caustic crossing, finite source, or parallax effects; \citealt{Wyrzykowski2020}; \citealt{Kaczmarek2022}, or via the astrometric signal; \citealt{Kains2017}; \citealt{Sahu2022}; \citealt{Lam2022}), provide enough information to break degeneracies between the lens mass, lens and source distance, and relative proper motion.

Instead of large-scale monitoring campaigns, microlensing events can also be found via prediction \citep{Refsdal1964}. If the proper motions and parallaxes of a large number of stars are known, future close alignments and hence microlensing events can be predicted. This permits targeted followup campaigns to be organized ahead of time to detect the microlensing signals. In this case, because the lens and source identity are known and we have lens and source astrometric information, degeneracies in both the photometric \citep{Paczynski1995, McGill2019a} and astrometric \citep{Paczynski1996} microlensing signal can be broken and the lens mass can be constrained \citep{Sahu2017, Zurlo2018, McGill2022}.

Since the suggestion of \cite{Refsdal1964} to use astrometric catalogs to predict microlensing events, it has been attempted many times \citep[e.g.,][]{Salim2000, Proft2011} with work in this area being reignited by the advent of high-precision astrometry from Gaia \citep[e.g.,][]{McGill2018, Kluter2018b, Bramich2018, Mustill2018}. While Gaia has proved a significant source of predicted microlensing events with over 5000 events being found \citep{McGill2020, Kluter2021}, Gaia lacks data in key areas. This motivated searches combining Gaia data with other catalogs to extend searches and find missing events. \cite{Ofek2018} used Gaia Data Release two \citep[GDR2;][]{GDR22018} as background sources combined with a pulsar proper motion catalog and found one promising candidate. \cite{Nielson2018} used faint low-mass objects as lenses from Pan-STARRS data release 1 \citep{Flewelling2020} missing from GDR2 to search for events, and found 27 candidate events occurring before the year 2070. Finally, \cite{McGill2019b} combined GDR2 with the data from the VISTA Variables in the Via Lactea survey \citep[VVV;][]{Minniti2010, Smith2018} to remedy Gaia’s incompleteness for background sources in the Galactic bulge and plane due to extinction and crowding and found two events missed by other searches.

Despite the efforts of previous predicted microlensing event searches, a search solely focusing on brown dwarfs lenses has yet to be carried out. The first naive approach to finding predicted brown dwarf events would be to focus on astrometric catalogs of the bulge where the microlensing optical depth (or probability of a microlensing event occurring) is high \citep{Dominik2000, Belokurov2002}. However, early searches for brown dwarfs in the WISE data were significantly biased against discovery in the Galactic plane due to source confusion in crowded fields (WISE band 1 and 2 pixels are $\sim$ 2.75$\arcsec$; \citealt{wright2010}) and higher infrared backgrounds. Although the more recent CatWISE 2020 processing significantly corrected for bias in the Galactic plane \citep{marocco2021}, finding thousands of new sources, \cite{kirkpatrick2021} found that the number of brown dwarfs in this area of the sky remains incomplete at $\sim 15 \%$. Moreover, the majority of brown dwarfs with precise astrometry and high proper motions lie outside the Galactic plane.

Therefore, in this paper, we search for predicted microlensing events using the two most recent and complete samples of brown dwarfs in the Solar Neighborhood, \citep{kirkpatrick2019,best}. We also search for objects using the UltracoolSheet \citep{ultracool_sheet}. To counter the issue of low optical depth outside the Galactic plane we use background sources from the Dark  Energy  Camera  Legacy  Survey  \citep[DECaLS;][]{dey},  which  reaches  a  5$\sigma$ depth of $g=23.95$ and is significantly deeper than Gaia \citep[G=20.7 depth;][]{Gaia}. Ultimately, the combination of these two datasets allows the search for predicted microlensing events by brown dwarfs missed by previous searches.

We focus our efforts on finding events with astrometric signals rather than photometric signals for two reasons. Firstly, the astrometric optical depth and size of the astrometric signals are large for our close-by sample of brown dwarfs \citep{Dominik2000, jordi}. Secondly, the astrometric signals are less sensitive to the predicted lens-source alignment than the photometric signals making it easier to predict astrometric signals with high confidence  \citep[see e.g.,][]{Bramich2018}, which is important given the limited precision of the astrometric catalogs we use for prediction.

Section \ref{background} gives a brief overview of the theoretical background for microlensing. Section \ref{data} presents the the brown dwarf and background star catalogs used in our search. Section \ref{methods} details our new method for finding predicted astrometric microlensing events. We calculate the rate of expected astrometric microlensing events for nearby brown dwarfs and we discuss how to increase this rate in the context of current and future astrometric catalogs in Section \ref{results}. Section \ref{conc} provides a summary and our recommendations to find future brown dwarf astrometric microlensing events. This paper presents publicly available code which may be found here\footnote{\url{https://github.com/JudahRockLuberto/mlfinder}}.

\section{Theoretical Background}\label{background}

\subsection{Astrometric microlensing}

In this section, we cover the required background for astrometric microlensing. For a detailed review of the topic, we defer to \cite{Dominik2000} and \cite{Bramich2018}. Consider a lens with mass $M_{L}$ at a distance $D_{L}$ intervening between an observer and background source (hereafter the source) at distance $D_{S}$, where  $D_{L}<D_{S}$. In the case of perfect lens-source alignment, gravitational lensing causes an image of the source in the shape of a circular Einstien ring to be formed with angular radius \citep{Chwolson1924,Einstein1936},

\begin{equation}
    \theta_{E} = \sqrt{\frac{4GM_{L}}{c^2}(D_{L}^{-1} - D_{S}^{-1})}\stackrel{D_{S}\gg D_{L}}{\approx}\sqrt{\frac{4GM_{L}}{D_{L}c^2}}.
\end{equation}

Here, $G$ is the gravitational constant and $c$ is the speed of light. In this paper, we use the close lens and distant source approximation ($D_{S}\gg D_{L}$) which is justified for our lens sample of brown dwarfs within $25$pc. Having nearby lenses also allows $D_{L}$ to be constrained by parallax ($1/D_{L}=\pi_{L}$; $\pi_{L}$ is the parallax of the lens).

In the case of imperfect lens-source alignment, a major (+) and minor (-) image of the source are formed with positions,

\begin{equation}
    \boldsymbol{\theta}_{\pm}(\boldsymbol{u}) = \left(u\pm\sqrt{u^{2}+4}\right)\frac{\Theta_{E}}{2}\hat{\boldsymbol{u}}.
    \label{eq:image_positions_vec}
\end{equation}

Here, $\boldsymbol{u} = (\boldsymbol{\phi}_{S}-\boldsymbol{\phi}_{L})/\theta_{E}=\boldsymbol{\beta}/\theta_{E}$ is the normalised lens-source angular separations vector. $\boldsymbol{\phi}_{S}$ and $\boldsymbol{\phi}_{L}$ and the angular positions of the source and lens, and $\boldsymbol{\beta}$ is the lens-source angular separation. $\hat{\boldsymbol{u}}$ and $u$ are the unit vector and magnitude of $\boldsymbol{u}$, respectively. 

During a microlensing event, the source images change position and brightness giving rise to an astrometric signal. If the lens-source separation is large ($u\gg1$) and the flux of the lens ($F_{L}$) is small compared with the flux of the source ($F_{S}$), the astrometric shift due to microlensing is approximately \citep{Bramich2018},

\begin{equation}
    \boldsymbol{\Delta\theta}\stackrel{\substack{u\gg1 \\ F_{S}\gg F_{L}}}
{\approx}\frac{\theta_{E}}{u}\hat{\boldsymbol{u}}.
    \label{eq:shift}
\end{equation}

For our brown dwarf lenses, we can assume that $F_{L}\ll F_{S}$ if we restrict astrometric follow-up to be in the optical wavelengths where brown dwarfs are faint or completely dark. The $u\gg1$ assumption is dependent on the geometry of each specific event, and its safety can be checked easily on a case-by-case basis. In regimes where either of the assumptions break down, Eq. (\ref{eq:shift}) overestimates the size of the astrometric microlensing signal. Overall, if the astrometric signal is measured for an event and the astrometric parameters (positions, proper motions and parallaxes) of the lens and source are known, $\theta_{E}$ can be measured and the lens mass can be constrained. 

\subsection{Predicting Astrometric Microlensing Events}

In previous studies, microlensing events are predicted by searching for lens-source alignments that give rise to a threshold microlensing signal magnitude \citep[e.g.,][]{Bramich2018, Mustill2018}. For this method, a lens-mass estimate is required to predict the size of the microlensing signal. Lens mass estimates are obtained by either a mass-luminosity relationship for lens stars on the main sequence \citep[e.g.,][]{Kluter2018b} or by fitting appropriate evolutionary models to broad-band photometry \citep[e.g.,][]{Bramich2018}. While these different lens mass estimation techniques agree for the most part, they can disagree particularly for low-mass objects. The differing lens mass estimates can change the size of predicted microlensing signals enough to influence followup strategies \citep{McGill2019a}.

If determining the mass of the lens is the goal of predicting and following up an astrometric microlensing event \citep[e.g.,][]{Kluter2020}, events that allow the lens mass to be determined with some target precision should be searched for instead of searching for events with some peak threshold astrometric signal. To find events meeting a lens mass target precision ($\delta M_{L}$) we use the results of \cite{cushing}.  Specifically, taking the derivative of the magnitude of Eq. (\ref{eq:shift}) with respect to $M_{L}$, and rearranging with $\beta = |\boldsymbol{\beta}|$ we obtain,

\begin{equation}
\delta M_{L} = \frac{c^{2}\delta\Delta\theta\beta}{4G\pi_{L}}. 
\label{eq:mass_precision}
\end{equation}

This equation allows the estimation of the precision at which a lens mass can be determined given a lens-source separation and an astrometric precision of $\delta\Delta\theta$. The power of using Eq. (\ref{eq:mass_precision}) to search for events comes from the fact that it does not depend on the mass of the lens. This is because Eq. (\ref{eq:shift}) is linear in the lens mass. Lens mass estimates are not required to search for predicted microlensing events when using this method. Instead, we set a lens mass precision goal to search for useful predicted events that will provide useful mass constraints for a given object.

\section{Data} \label{data}

\subsection{Background Source Catalog}

We use the DESI Legacy Survey Data Release 9\footnote{\url{https://datalab.noirlab.edu/ls/ls.php}} (DR9) as our background source catalog, which has astrometry tied to the GDR2 reference frame. The DESI Legacy Imaging Surveys \citep[Legacy Survey;][]{dey} are a group of surveys which include the Dark Energy Camera Legacy Survey (DECaLS), the Mayall z-band Legacy Survey (MzLS) \citep[Legacy Survey;][]{dey}, and the Beijing-Arizona Sky Survey \citep[BASS;][]{zhou2017}. DECaLS is a $\sim14800$ deg$^2$ survey in three optical bands ($g$, $r$, and $z$) using the Dark Energy Camera \citep{flaugher} on the 4m Blanco telescope at the Cerro Tololo Inter-American Observatory. DECaLS data reach 5$\sigma$ depths of $g=23.95$, $r=23.54$, and $z=22.50$ in AB mag for point sources in individual images \citep{dey}. 

\begin{figure*}
    \centering
    \includegraphics[width=110mm]{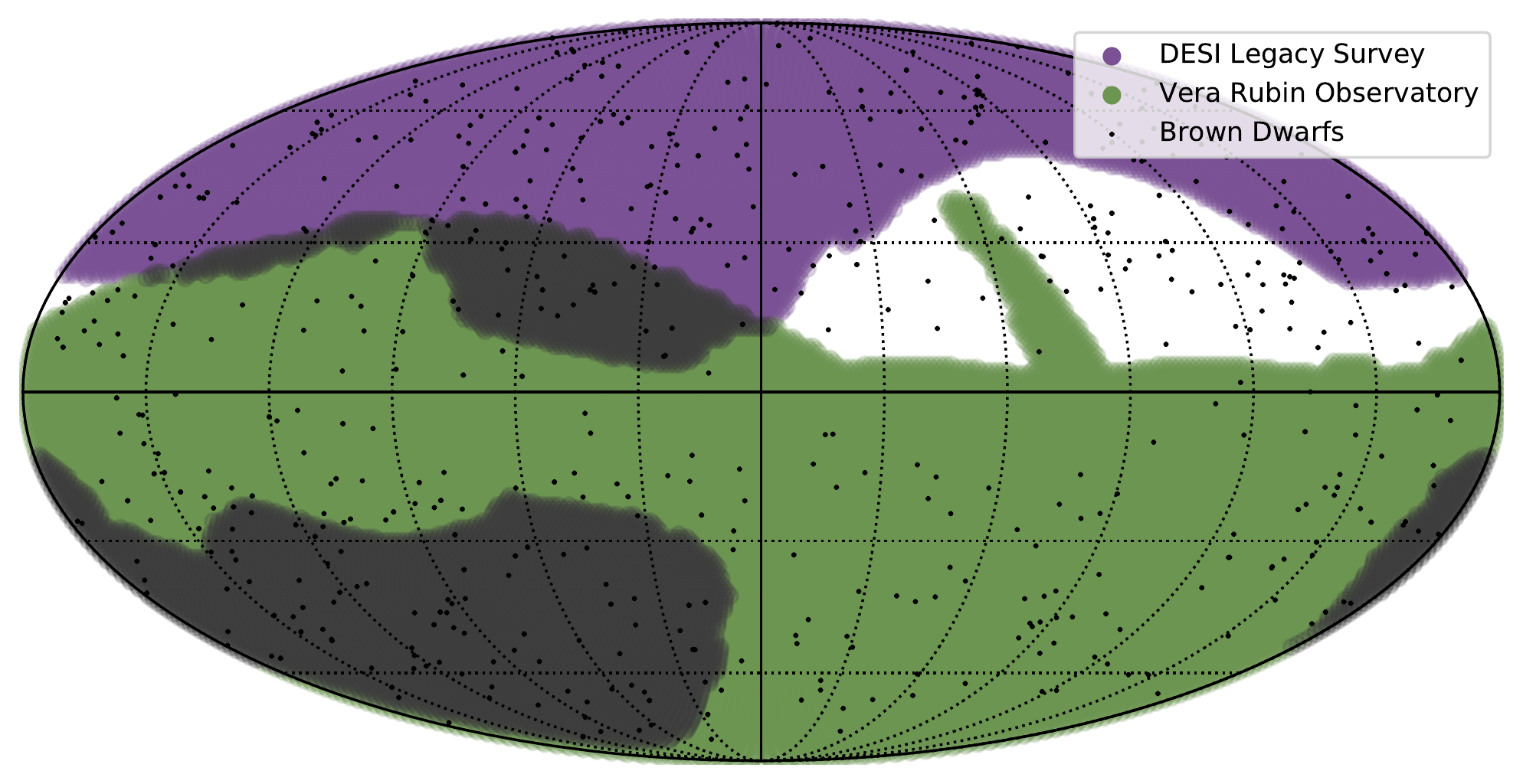}
    \caption{Sky coverage in Galactic coordinates of the Legacy Survey, the upcoming Legacy Survey of Space and Time by the Vera C. Rubin Observatory (VRO), as well as the positions of the brown dwarfs from \cite{kirkpatrick2021} and \cite{best} (black dots). The Legacy Survey is in purple, VRO is in green, and the overlap is in grey. The Legacy Survey (used in this paper) covers most of the northern hemisphere, while VRO covers the southern hemisphere and the Galactic plane.}
    \label{fig:mollweide}
\end{figure*}

We used the Legacy Survey `type' flag to select stellar objects. We made a magnitude cut to the star catalog to select sources brighter than $g<23.95$. This cut also removes background stars without measured magnitudes. The average magnitude of stars in the Legacy Survey catalog is 21.54 in g-band. However, even though the Legacy Survey detects more stars per unit area than \textit{Gaia}, the Legacy Survey does not cover as much sky. Figure  \ref{fig:mollweide} shows the footprints of the Legacy Survey and the upcoming Legacy Survey of Space and Time by the Vera C. Rubin Observatory \citep{ivezic2019}.

\subsection{Brown Dwarf Lenses}

We selected 1905 unique objects from \cite{kirkpatrick2021}, \cite{best}, and \cite{ultracool_sheet}. 314 of these were brown dwarfs from \cite{kirkpatrick2021}, a volume-limited sample of T and Y dwarfs within 20 pc. Multi-epoch \emph{Spitzer} data were used to fit their proper motion, $\mu_{\alpha}$ and $\mu_{\delta}$, and parallax, $\pi$, with parallactic uncertainties $\leq$10\% . We cut out targets that were noted to have poor quality astrometric solutions. 

An additional 231 objects originate uniquely from the \cite{best} catalog of L0-T8 dwarfs. The \cite{best} catalog combines measurements from UKIRT/WFCAM with existing data from the literature to produce a volume-limited survey out to 25 pc. We limit our selection from this catalog to brown dwarfs with parallax errors $\leq$10\% and astrometric uncertainties $\leq$12 mas per epoch.

We also select 1360 targets from the UltracoolSheet \citep{ultracool_sheet}, limiting our selection to objects of spectral type M6 or later that are not classified as exoplanets and have known parallax.

 A total of 1225 unique objects fall within the Legacy Survey area, marked as black points in Figure~\ref{fig:mollweide}, which we consider further for predicted astrometric microlensing.

\section{Methodology} \label{methods}

\subsection{Initial lens-source matching}

To narrow our search, for each brown dwarf lens, we select all background sources that are within 5 arcminutes of its reference position. We used a large selection radius because our lens sample contains fast moving brown dwarfs, (e.g., WISE J085510.83-071442.5 has proper motion of $8151.55 \pm 1.3$ mas/yr). 

The Dark Energy Survey (DES) imaged thousands of brown dwarfs \citep{rosell} in the red-optical. Thus, we found that some of our selected brown dwarfs were also identified in the Legacy Survey star catalog. Since we are only interested in background stars, and to avoid finding spurious events \citep{McGill2020}, it is desirable to remove the target brown dwarfs from the Legacy Survey star catalog. We found that the $g<23.95$ successfully removed brown dwarfs from our background star catalog. The reason for this is that brown dwarfs are generally red \citep{kirkpatrick2005}. The nearest L and T dwarfs can have optical magnitudes bright enough to be found in \textit{Gaia}, with typical \textit{Gaia G} - 2MASS \textit{J} colors of 4.5 and typical \textit{G} magnitudes fainter than $G>18$ \citep{smart2017}. Of the 52710 background stars selected, 37522 had dereddened g-band magnitudes dimmer than 20.7 mag (the \textit{Gaia} point source limit) demonstrating the power of deep photometric surveys for providing large catalogs of background stars.

\subsection{Lens-Source Trajectories} \label{trajectories}

In order to predict the lens-source angular separation ($\boldsymbol{\beta}$) we need to predict the lens and source angular positions. We computed trajectories across the celestial sphere using the \citet{smartgreen} equations:
\begin{multline}
\boldsymbol{\phi}(t) =
 \begin{bmatrix}
 \alpha_{\text{ref}} \\
 \delta_{\text{ref}}
 \end{bmatrix}
 + (t-t_{\text{ref}})
 \begin{bmatrix}
     \mu_{\alpha*}/\cos\delta_{\text{ref}} \\
     \mu_{\delta}
\end{bmatrix} \\ +\pi \boldsymbol{J}^{-1}\boldsymbol{R}_{\oplus}(t).
\label{eq:celestial_motion}
\end{multline}

Here, $(\alpha_{\text{ref}},\delta_{\text{ref}})$ is the reference position at time $t_{\text{ref}}$. $\mu_{\alpha*}$ and $\mu_{\delta}$ are the tangent plane proper motions in the directions of Right Ascentions and declination, respectively. $\pi$ is the parallax. $\boldsymbol{R}_{\oplus}(t)$ is the barcentric position of the Earth and was retrieved from JPL's Horizons Online Ephemeris System\footnote{\url{https://ssd.jpl.nasa.gov/horizons/}}. $\boldsymbol{J}^{-1}$ is the inverse Jacobian matrix of the transformation from Cartesian to spherical coordinates at the reference position.

For computation of $\boldsymbol{\phi}_{L}$ we used Eq. (\ref{eq:celestial_motion}) because all of the lens astrometric quantities were available. For the background sources, proper motions and parallax were not available from the Legacy Survey. Therefore we assumed that the source was distant enough to have negligible proper motion and parallax and $\boldsymbol{\phi}_{S}$ to be fixed at the Legacy Survey reference position. In the case that the background source was in GDR2, we used its GDR2 astrometric parameters to calculate its trajectory using Eq. (\ref{eq:celestial_motion}). 

\subsection{Checking for Microlensing Events}

With the brown dwarf lenses and the catalog of surrounding background sources in hand, we checked for microlensing events. We computed trajectories across the geocentric celestial sphere of the source and lens in one-day increments. Then, for each  lens-source pair, and assuming an astrometric precision of $\delta\Delta\theta=0.2$ mas \citep[inline with current Hubble Space Telescope capabilities;][]{Kains2017, Sahu2017}, we computed $\delta M_{L}$ using Eq. (\ref{eq:mass_precision}) at each increment. Finally, we recorded the minimum $\delta M_{L}$ across all increments for each lens-source pair. If the minimum $\delta M_{L}$ was $<1000 M_{\text{jup}}$, we investigated the event further.

For the surviving events, we performed a Monte Carlo simulation to account for the uncertainty in the lens-source astrometric parameters. Specifically, we drew 500 samples from appropriate Gaussian distributions for the lens reported positions, proper motions, and parallaxes. For the background sources, we pulled the reported uncertainties from GDR2. In the case of background sources from the Legacy survey without an astrometric solution, we approximated them as motionless with respect to GDR2 and with a fixed positional uncertainty on the order similar to other background stars with astrometric solutions, or $\sim 10^{-6}$ deg. For each of the samples we repeated the procedure to find the minimum $\delta M_{L}$, which gave a distribution of minimum $\delta M_{L}$ for each lens-source pair. For our final sample we only consider events where the lens is a known brown dwarf.

\section{Results} \label{results}

\subsection{Microlensing Event Search}

We searched for candidate microlensing events occurring between the years 2014 and 2032. Astrometric coverage from \textit{Gaia} began in 2014, and it is possible that past events are detectable in the \textit{Gaia} data. 

\begin{figure}
    \centering
    \includegraphics[width=8cm]{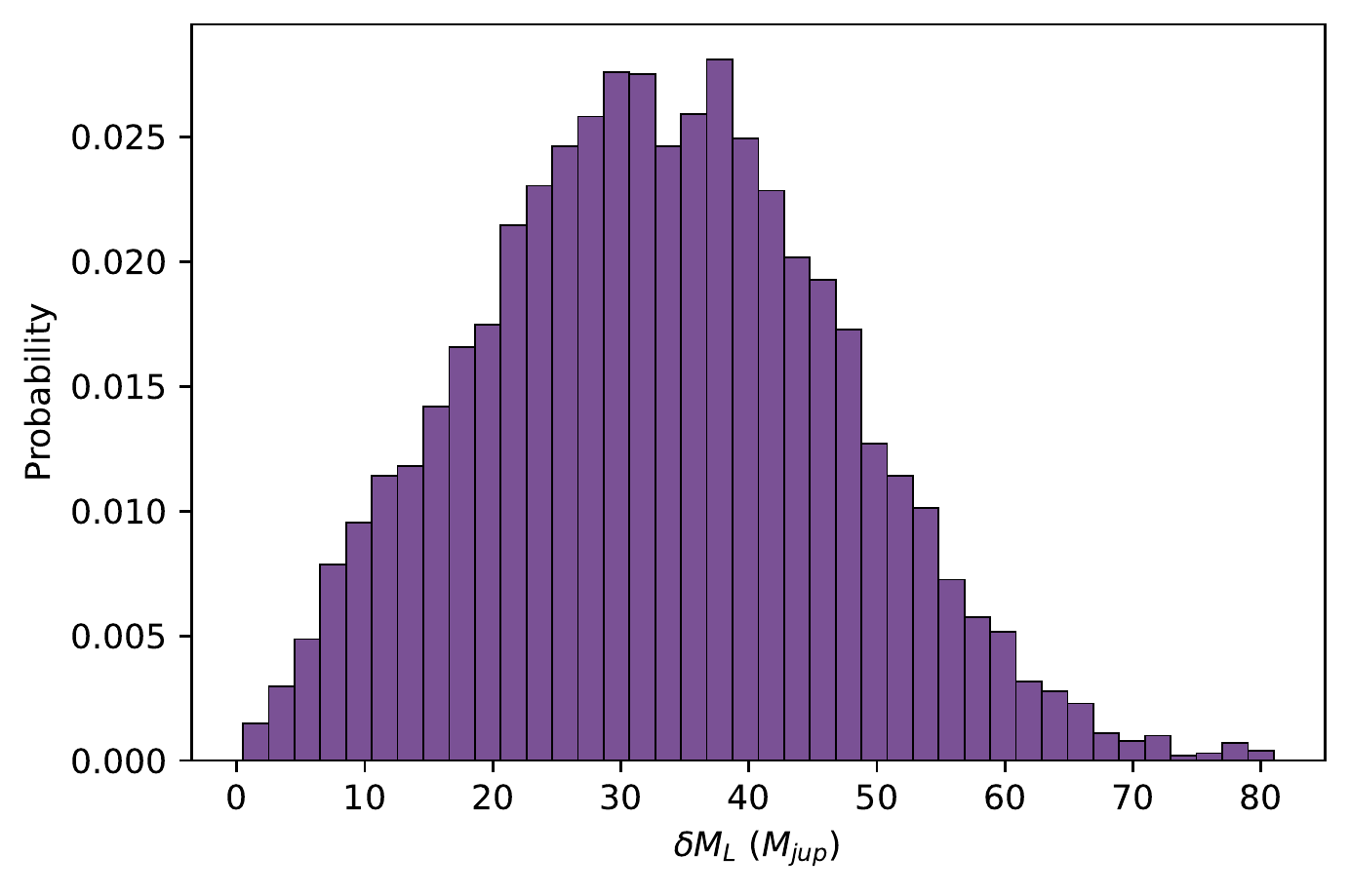}
    \caption{The mass uncertainty of an astrometric microlensing event of WISE 0413-4750 as predicted using a Monte Carlo simulation. 5000 samples were used, using the uncertainty of the motions and positions of the brown dwarf and the background star. The mass uncertainty is $\delta M_{L} = 32.04\pm14.76 M_{\text{jup}}$.}
    \label{fig:0413_massuncertainty}
\end{figure}

Our search yielded no astrometric microlensing candidate events with a minimum mass uncertainty $\sim10M_{\text{jup}}$, and of the sample, only 35 had a $\delta M_{L} < 1000 M_{\text{jup}}$. These results are tabulated in Table \ref{tab:all_objects}. The future event with the smallest mass uncertainty was caused by WISE J041358.14-475039.3, hereafter WISE 0413-4750, with an expected minimum $\delta M_{L} = 32.04 \pm 14.7 \ M_{\text{jup}}$. A distribution of $\delta M_{L}$ is shown in Figure \ref{fig:0413_massuncertainty}, created using a Monte Carlo of 5000 samples. $7.8 \%$ of the samples resulted in $\delta M_{L} < 10 M_{\text{jup}}$. This event has a minimum predicted lens-source separation of $63.16 \pm 29.1$ mas, and a time of minimum separation in late-December, 2031 ($2031.8453 \pm 0.2811$ julian year). 

\begin{deluxetable*}{c|cccccc}
\tabletypesize{\footnotesize}
\tablecolumns{6}
\tablewidth{0pt}
\tablecaption{Brown Dwarf microlensing event parameters\label{tab:all_objects}}
\tablehead{
    \colhead{Object Name} & \colhead{Ref.} & \colhead{$J_{\text{MKO}}$} & \colhead{$g_{\text{bs}}$} &  \colhead{$t_{\text{min}}$} & \colhead{$\theta_{\text{min}}$} & \colhead{Pred. Mass Unc.} \\ 
    \colhead{} & \colhead{} & \colhead{(mag)} & \colhead{(mag)} & \colhead{(yrs)} & \colhead{(mas)} & \colhead{($\mathrm{M_{jup}}$)}
    }
\colnumbers
\startdata
    WISE J004945.61+215120.0 & K21 & 16.36$\pm$0.02 & 19.99 & 2023.7871$\pm$0.0014 & 3010.16$\pm$7.88 & 551.44$\pm$1.44\\
    WISEPA J020625.26+264023.6 & UCS & 16.42$\pm$0.11 & 23.16 & 2022.7868$\pm$0.0025 & 1292.26$\pm$4.29 & 637.96$\pm$2.12\\
    2MASSW J0228110+253738 & B20 & 13.76$\pm$0.02 & 23.35 & 2018.4783$\pm$0.1091 & 161.64$\pm$10.44 & 150.63$\pm$9.73\\
    PSO J043.5395+02.3995 & UCS & 15.92$\pm$0.01 & 13.75 & 2025.3894$\pm$0.0005 & 3034.24$\pm$5.87 & 534.17$\pm$1.03\\
    WISE J041358.14-475039.3 & K21 & 19.71$\pm$0.13 & 23.47 & 2031.8453$\pm$0.2811 & 63.16$\pm$29.1 & 32.04$\pm$14.76\\
    2MASSI J0428510-225323 & UCS & 13.47$\pm$0.02 & 22.48 & 2015.6787$\pm$0.0026 & 133.72$\pm$2.6 & 90.75$\pm$1.76\\
    2MASSW J0832045-012835 & B20 & 14.08$\pm$0.02 & 23.41 & 2017.7333$\pm$0.2935 & 392.39$\pm$15.24 & 238.59$\pm$9.27\\
    WISE J083337.83+005214.2 & K21 & 20.28$\pm$0.1 & 22.42 & 2023.8163$\pm$0.0023 & 2871.42$\pm$14.0 & 926.65$\pm$4.52\\
    WISE J085510.83-071442.5 & K21 & 25.0$\pm$0.5 & 21.43 & 2016.3127$\pm$0.0005 & 10451.48$\pm$2.3 & 612.34$\pm$0.13\\
    SDSS J111316.95-000246.6 & UCS & 15.01$\pm$0.01 & 23.74 & 2016.9645$\pm$0.002 & 145.03$\pm$6.11 & 162.89$\pm$6.86 \\
    DENIS-P J1159+0057 & B20 & 14.02$\pm$0.02 & 23.22 & 2023.4572$\pm$0.005 & 310.45$\pm$46.33 & 234.85$\pm$35.05 \\
    WISEPC J131106.24+012252.4 & UCS & 18.97$\pm$0.08 & 20.82 & 2028.0405$\pm$0.0019 & 858.68$\pm$6.53 & 321.01$\pm$2.44 \\
    2MASSW J1411175+393636 & B20 & 14.55$\pm$0.02 & 23.95 & 2015.8066$\pm$0.002 & 197.51$\pm$3.06 & 172.79$\pm$2.68 \\
    SDSS J161928.31+005011.9 & B20 & 14.35$\pm$0.02 & 23.18 & 2020.5945$\pm$0.0067 & 494.5$\pm$18.38 & 356.27$\pm$13.24 \\
    2MASSI J1807159+501531 & UCS & 12.88$\pm$0.02 & 22.17 & 2015.1479$\pm$0.0025 & 42.57$\pm$1.52 & 16.02$\pm$0.57 \\
    SDSS J202820.32+005226.5 & B20 & 14.18$\pm$0.02 & 18.18 & 2024.7464$\pm$0.0066 & 1048.75$\pm$20.57 & 878.65$\pm$17.23 \\
    WISE J220905.73+271143.9 & K21 & 22.86$\pm$0.13 & 13.6 & 2022.503$\pm$0.0009 & 3418.94$\pm$6.26 & 543.83$\pm$1.0 \\
    WISE J222219.93+302601.4 & B20 & 16.58$\pm$0.02 & 17.17 & 2026.9971$\pm$0.0479 & 682.32$\pm$26.25 & 635.86$\pm$24.46 \\
    HIP 112422B & UCS & 16.02$\pm$0.02 & 23.11 & 2020.0695$\pm$0.0279 & 60.15$\pm$5.15 & 101.12$\pm$8.66
\enddata

\tablecomments{Objects with a predicted microlensing event of mass uncertainty $<1000 M_{\text{jup}}$, sorted by right ascension and declination. The astrometric information of each brown dwarf can be found using the reference column, where K21 is \cite{kirkpatrick2021}, B20 is \cite{best}, and UCS is \cite{ultracool_sheet}. All astrometry was performed in the celestial reference system (ICRS). The J-band MKO magnitude is in the $\mathrm{J_{MKO}}$ column, where objects without a given magnitude are denoted by a slash. The g-band magnitude of the background star, as given by the Legacy Survey, is in the $\mathrm{g_{bs}}$ column. $\theta_{\text{min}}$ and the predicted mass uncertainty are given as median values. In the event one object had multiple events with predicted $\delta M_{L} < 1000 M_{\text{jup}}$, the event with the smallest predicted mass uncertainty was chosen. \textit{Table \ref{tab:all_objects} is published in its entirety in the machine readable format at \url{https://github.com/JudahRockLuberto/mlfinder.}}}
\end{deluxetable*}

\begin{figure}
    \centering
    \includegraphics[trim={11.5cm 0 12cm 0},clip, width=8cm]{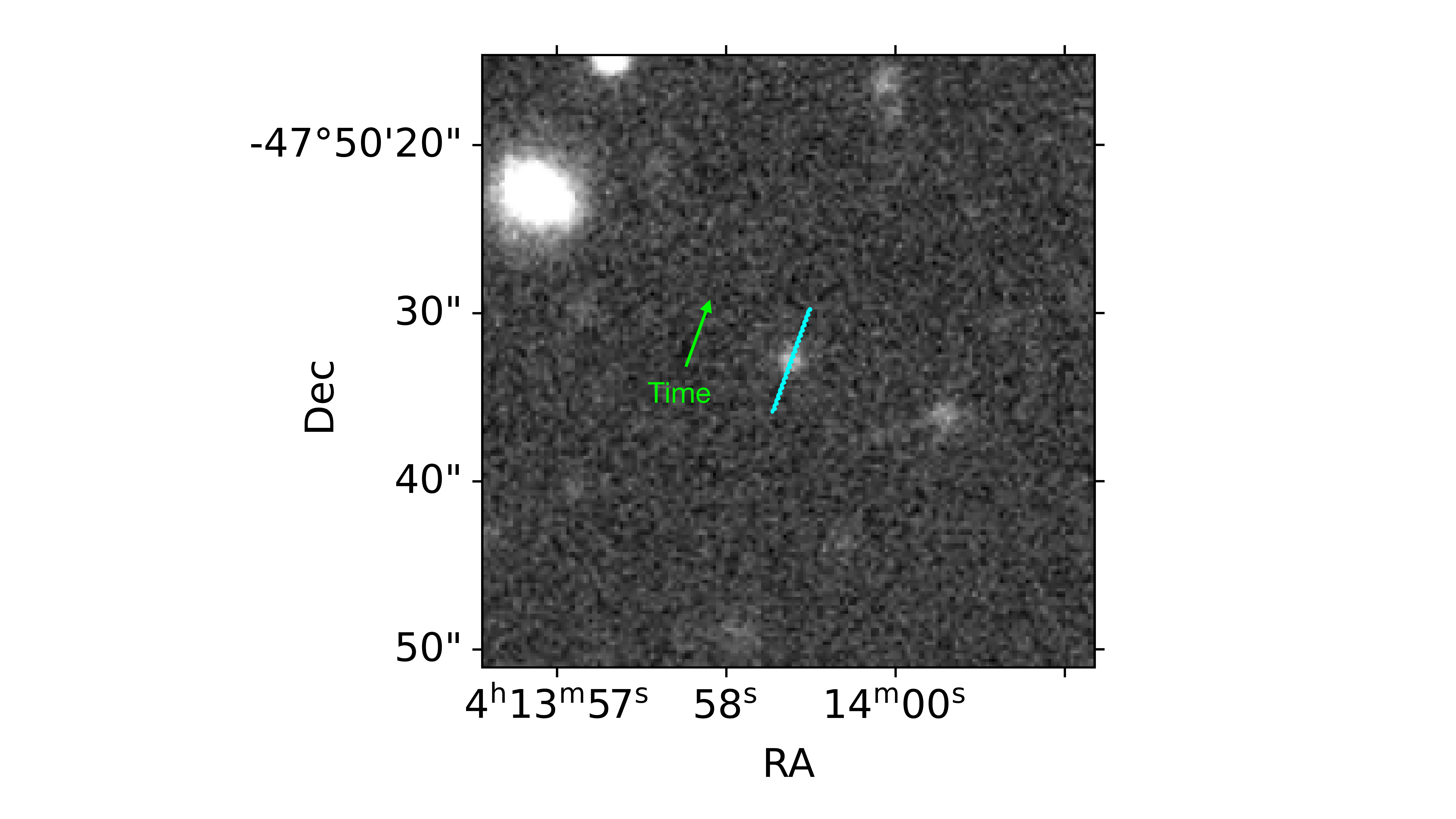}
    \caption{The path of WISE 0413-4750 from 2022 - 2042 plotted over a Legacy Survey data release 9 image. The cyan line is the astrometric path of the brown dwarf and the green arrow near the brown dwarf's path indicates the direction of the brown dwarf, beginning in the bottom left and moving towards the top right. The brown dwarf has a minimum separation in December 2031 of $0.07 \pm 0.03$ arcseconds. While no measurable astrometric microlensing event occurs in this system, it is useful to note that this background star is not found in GDR2, demonstrating the power of deep photometric surveys for providing additional background sources.} 
    \label{fig:path}
\end{figure}

WISE 0413-4750's path is plotted on top of the Legacy Survey DR9 image in Figure \ref{fig:path}. The cyan line is the path of the brown dwarf, and the green arrow shows the brown dwarf's direction of motion. The background star (\textit{g} mag = 23.47) is the star under the brown dwarf path. The background source is too faint to be found in GDR2.

We also searched for useful mass constraints using the set of predicted events by very low-mass stars from \cite{Nielson2018}. We analyzed each event in \cite{Nielson2018} using the lens astrometry they reported and our event finding method. Out of the 27 events found in \cite{Nielson2018}, only 2 occurred in the Legacy Survey coverage. Encouragingly, using background source astrometry from the Legacy Survey, we recovered both events with similar events parameters to \cite{Nielson2018}. However, the predicted mass constraints were greater than 75 $M_{\text{jup}}$, and thus too large to place useful constraints on the lens mass. Using the lens and background source astrometry from \cite{Nielson2018}, of the 27 events, 22, 12, and 6 events have predicted average mass uncertainties of $<1000$M$_{\text{jup}}$, $<100$M$_{\text{jup}}$, and $<50$M$_{\text{jup}}$, respectively. We include a table of all the predicted mass uncertainties for the \cite{Nielson2018} sample of events at the GitHub repository associated with this paper.\footnote{\url{https://github.com/JudahRockLuberto/mlfinder.}}

\subsection{Expected Rates of Astrometric Microlensing Events}

\begin{figure*}
    \centering
    \includegraphics[width=\textwidth]{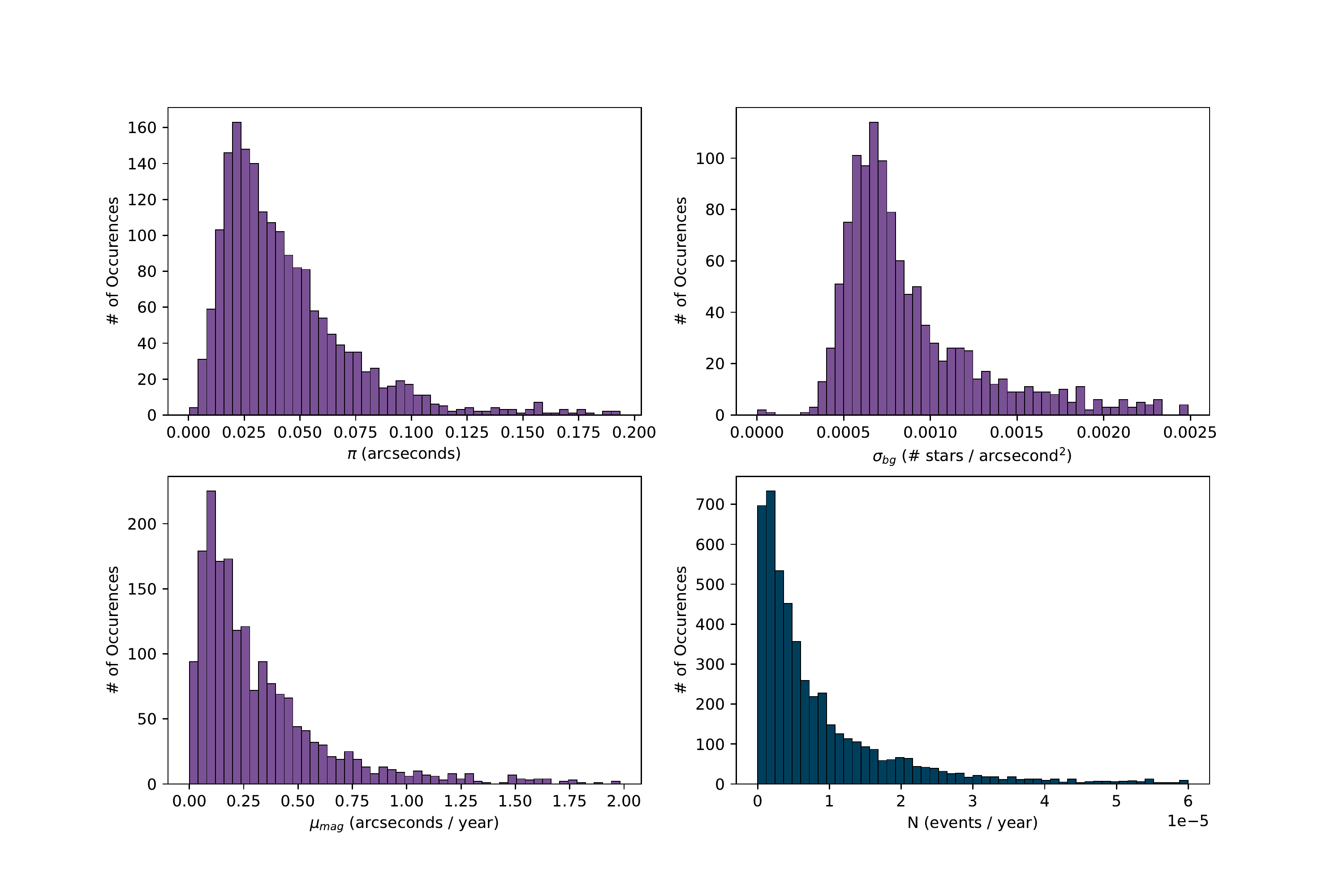}
    \caption{The distribution of parallaxes (upper-left), $\pi$, magnitudes of proper motion (upper-right), $\mu_{mag}$, and stellar densities (lower-left), $\sigma_{bg}$ for our sample of 1905 objects. This distribution is used in the random sampling of the motion of an average hypothetical brown dwarf to calculate the probability of an event per year for a hypothetical brown dwarf. Brown dwarfs with high parallax, proper motion, or star density were cut from the plot to show significant data. For the star densities, we computed each data point using a 5 arcminute radius circle. The number of events per year (bottom-right) is calculated for 5000 hypothetical brown dwarfs. For each brown dwarf, $\pi$, $\mu_{mag}$, and $\sigma_{bg}$ were randomly sampled from the known distributions of each observable. This plot is used in comparison to \cite{cushing}, who found the probability of events per year for a brown dwarf in the Galactic plane to be near 0.05, instead of our calculated distribution's trend of $< 0.0001$ events per year. The data above 0.00006 events / year were cut to plot only significant data.}
    \label{fig:simulatednumberofevents}
\end{figure*}

Here we discuss why few predicted astrometric microlensing events were found with this sample of brown dwarf lenses and background sources. We begin by computing the expected number of astrometric microlensing events by brown dwarfs in the Solar Neighborhood, per year. Following from Eq. (\ref{eq:mass_precision}) the rate of events derived by \cite{cushing} is,
\begin{equation}
    N = 2k\pi_{L}\mu_{L}\sigma_{bg}\left(\frac{\delta M_{L}}{\delta \Delta \theta}\right)
    \label{eq:numberofevents}
\end{equation}
Here,  $k=(4G/c^{2} \text{AU})\approx 8.14$ mas  $/M_{\odot}$. $\mu_{L}$ is the magnitude of the proper motion in arcseconds per year and $\sigma_{bg}$ is the surface density of stars in stars per square arcsecond. $\delta M_{L}$ and $\delta\Delta\theta$ are in units of $M_{\odot}$ and mas, respectively.  We generate a set of 5000 random L-Y brown dwarfs within the Solar Neighborhood and the Legacy Survey footprint, with values of $\pi_{L}$ and $\mu_{L}$ from our real and representative sample of brown dwarfs (Figure \ref{fig:simulatednumberofevents}).  We then compute $\sigma_{bg}$ by finding the distribution of surface density of stars within 5 arcminutes of all of our brown dwarfs and randomly sample from the $\sigma_{bg}$ distribution 5000 times (Figure \ref{fig:simulatednumberofevents}). Assuming a mass precision of $10 M_{\text{jup}}$, and an astrometric precision of $0.2$ mas, we compute the expected rate of events using Eq. (\ref{eq:numberofevents}).

Figure \ref{fig:simulatednumberofevents} shows the expected rate of events from our simulations. The maximum rate of events was $\sim 0.0004$ yr$^{-1}$ and the average was $\sim 10^{-5}$ yr$^{-1}$. The integrated rate over the distribution is $\sim$ 0.05 yr$^{-1}$, or around one event every twenty years. Because this simulation used 5000 hypothetical nearby, fast-moving brown dwarfs ($\sim$ 30 times the number used in this paper), we can conclude that an increased sample of brown dwarfs in the background field of the Legacy Survey is unlikely to increase the expected rate of events enough to give a reasonable chance of finding one event over the time scale we searched.  

Implicit in Eqs. (\ref{eq:mass_precision} and \ref{eq:numberofevents}) is the assumption that we are only taking one measurement of the astrometric microlensing signal at its maximum. Of course, an astrometric follow up campaign could collect many more data points during an event. Such a campaign would increase the precision we could measure $M_{L}$, so Eq. (\ref{eq:numberofevents}) is a lower limit on the event rate.  Typical astrometric followup campaigns with the Hubble Space Telescope (HST) have taken $\sim50-100$ exposures over $\sim10$ epochs for a predicted astrometric microlensing event \citep[e.g.,][]{Sahu2017, Sahu2019}. Assuming a $\sim1/\sqrt{N}$ mass precision increase, a HST campaign is unlikely to increase the lens mass precision by a factor $>10$. To summarize, the density of stars and brown dwarfs are too low in the regions of the sky considered in this paper to have any significant probability of finding an event with a scientifically useful ($\delta M_{L}\sim10 M_{\text{jup}}$) mass constraint.

\subsection{Future Prospects} \label{lsgaia}
We now turn to assessing future prospects for finding astrometric microlensing events by brown dwarfs. Given future observational facilities and their surveys, we examine whether the rate of events in Eq. \ref{eq:numberofevents} will improve significantly. First, the average $\pi_L$ and $\mu_L$ are not expected to change significantly. The current sample of known brown dwarfs in the Solar Neighborhood is mostly complete, although $\sim$ 15\% nearby brown dwarfs are likely missing in the Galactic Plane \citep{kirkpatrick2021}. However, these targets, when found, are unlikely to have a different distribution of $\pi_L$ and $\mu_L$ from the sample used here (see Fig. \ref{fig:simulatednumberofevents}).

The next factor that could be increased by future surveys is the background source density, $\sigma_{bg}$. There are two ways to increase $\sigma_{bg}$: take deeper observations, or observe in a higher density field like the Galactic Bulge or Galactic Plane. The former method is what we attempted in this paper, using the Legacy Survey. DECaLS, which is in the Legacy Survey, has a $5 \sigma$ point source depth of $g = 23.95$, where \textit{Gaia} has a point source threshold of $G = 20.7$. Using WISE 0413-4750 as the example brown dwarf in our sample again, only $\sim$5.7\% of the background stars from the Legacy Survey DR9 for the field around WISE 0413-4750 are also detected \textit{Gaia} point sources. Optical surveys can separate stars from galaxies to roughly 25th magnitude \citep[e.g.,][]{leauthaud2007}. To estimate the increase in stellar density from a star selection extending to $g=25$, rather than $g=23.95$, we use the COSMOS ACS catalog \citep[][]{leauthaud2007}. We find that going to $g=25$ would increase the stellar density by about a factor of 1.29. 

Although deeper surveys increase the background source density, there is a trade-off with astrometric precision, $\Delta \theta$, because astrometric precision is dependent on source brightness. For targeted follow-up campaigns using e.g. HST \citep[e.g.,][]{Sahu2017} or the upcoming Roman Space Telescope, astrometric precision on faint sources can be achieved with longer exposure times. For survey-mode imaging, such as with the upcoming Legacy Survey of Space and Time by the Vera C. Rubin Observatory \citep{ivezic2019} or the Microlensing Survey with Roman Space Telescope \citep{Penny2019}, astrometric precision on faint background stars may be a limiting factor. Figure \ref{fig:delta_ml_vs_massprecision} shows that even with an increased mass precision, to achieve a highly constrained brown dwarf mass measurement, we need brown dwarfs to pass with $\sim 50$ mas of a background source, which is most likely to occur in higher density fields like the galactic bulge.

The latter method for obtaining a higher $\sigma_{bg}$ was investigated in \cite{cushing}, with one hypothetical brown dwarf at $\ell = 45^{\circ}, b = 0^{\circ}$ near the Galactic Plane ($\sigma_{bg}$=0.12 stars/arcsec$^2$) and detectable by HST. They found the rate of astrometric microlensing events to be N $\approx$ 0.05 events/year for this object. \cite{cushing} used a mass precision of $0.10 M_{\odot}$ ($\approx 100 M_{\text{jup}}$), which is roughly $20$ times our value. However, this effect is negligible compared to the change in $N$ from the difference in the stellar density, which is higher by a factor of several hundred in the Galactic Plane, compared to the fields covered by the Legacy Survey.

\subsection{Public Release of Code for Future Searches} 

\begin{figure}
    \centering
    \includegraphics[width=8cm]{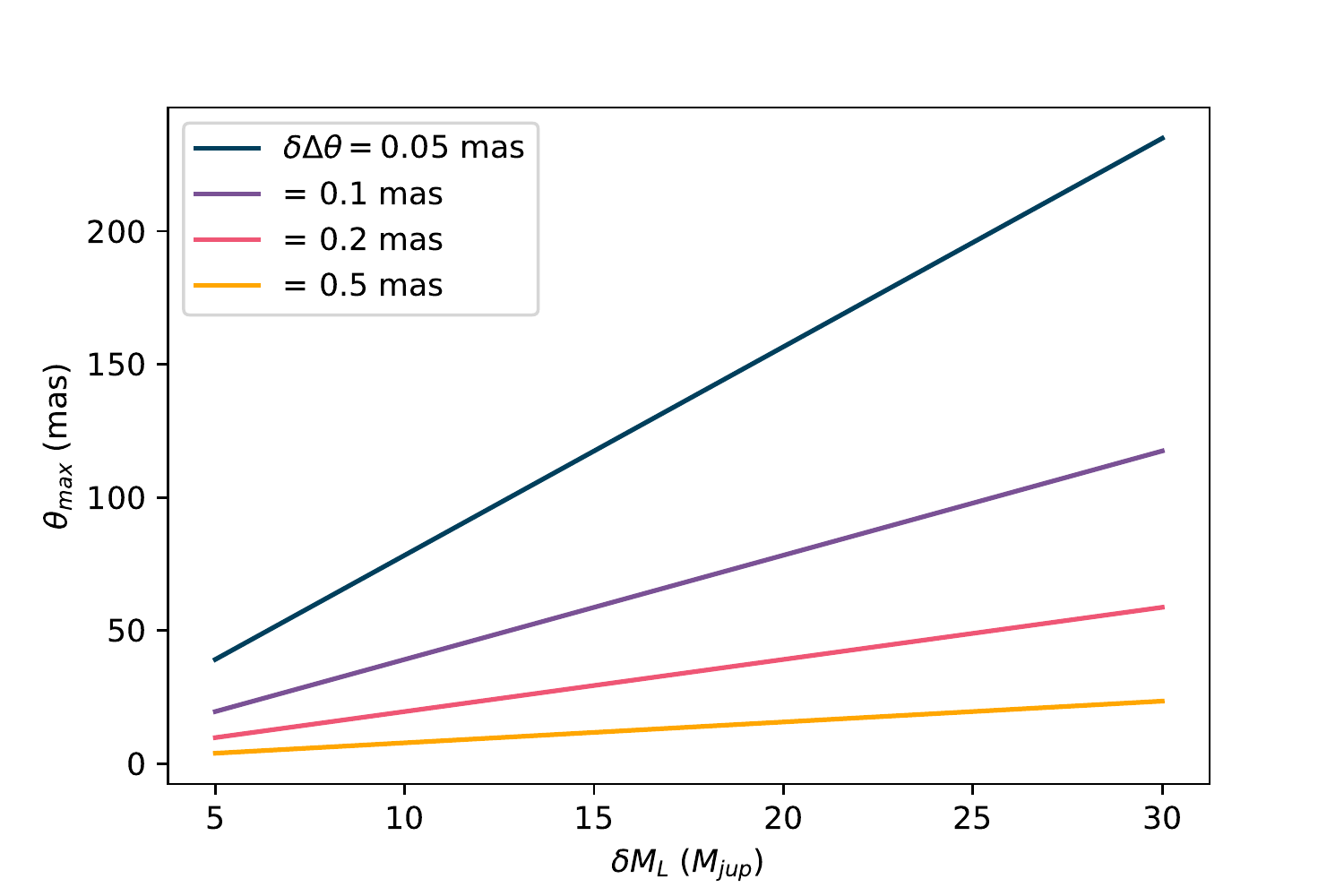}
    
    \caption{The maximum separation in milliarcseconds required for an event to yield a given mass uncertainty in $M_{\text{Jup}}$ using Eq. (\ref{eq:mass_precision}). This is calculated for a brown dwarf with a median parallax from our sample (50.35 mas). Several astrometric uncertainties are plotted, $\delta \Delta \theta = 0.2$ mas is inline with current HST capabilities \citep[e.g.,][]{Kains2017}. $\delta \Delta \theta = 0.05$ mas represents the effective precision a large observing campaign may achieve by taking multiple epoch of data \citep[][]{Sahu2017,Zurlo2018} or future space-based astrometric missions \citep[e.g.,][]{WFIRST2019}.}
    
    \label{fig:delta_ml_vs_massprecision}
\end{figure}

\begin{figure}
    \centering
    \includegraphics[width=8cm]{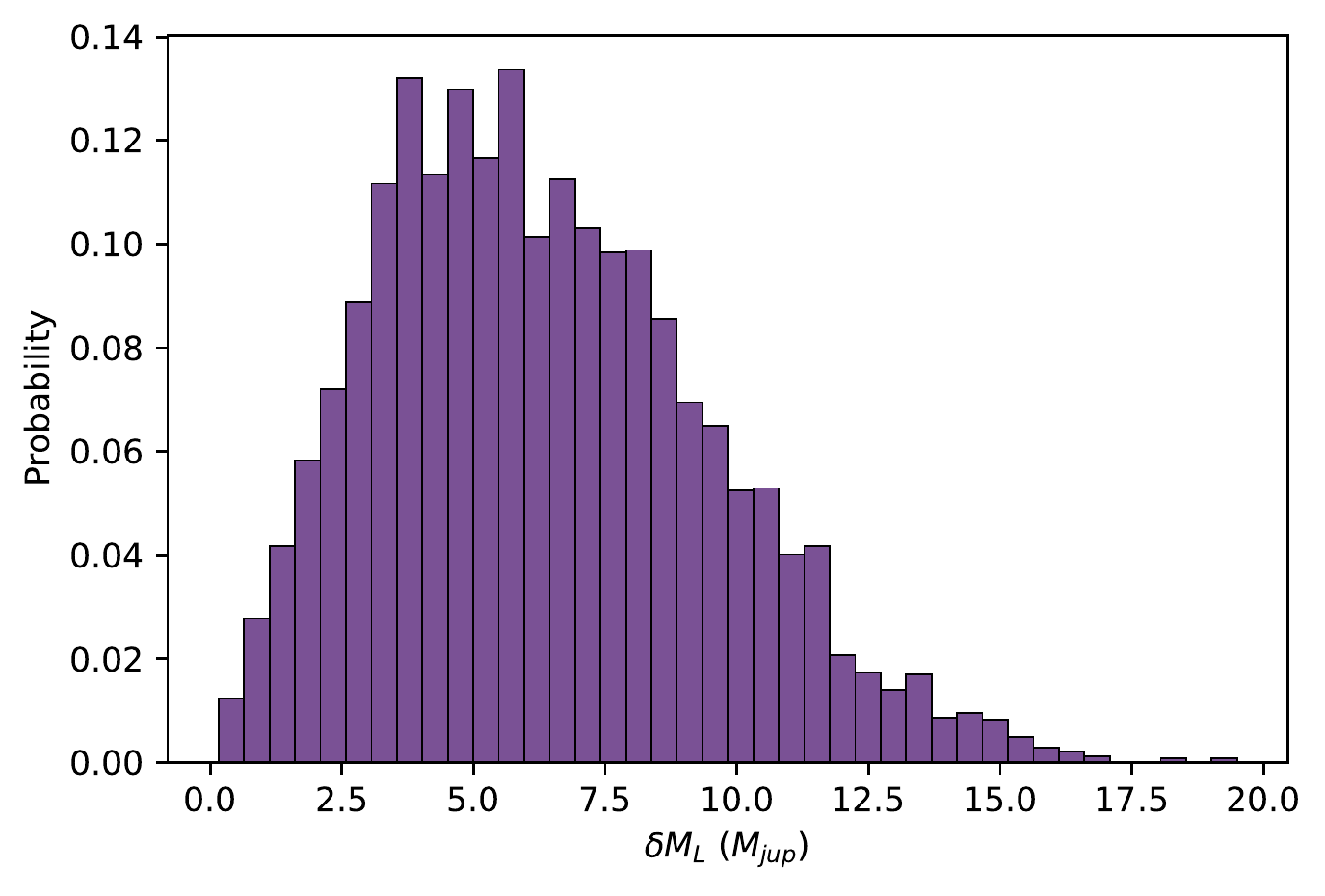}
    \caption{Distribution of $\delta M_{\rm{L}}$ calculated by using a Monte Carlo sampling of $\alpha$, $\delta$, $\pi$, $\mu_{\alpha}$, and $\mu{\delta}$. The motion and respective uncertainties were the average brown dwarf motions used in this paper. The minimum separation was 12 mas and the mass uncertainty calculated in this fashion is $\delta\Delta M_{L}=5.88\pm 3.28 M_{\text{jup}}$.}
    \label{fig:mcmc}
\end{figure}

Future searches of predicted microlensing events, such as those in the galactic plane, can be performed using our publicly available code\footnote{\url{https://github.com/JudahRockLuberto/mlfinder}}. This is the first publicly available code to search for predicted microlensing events without requiring a lens mass estimate. All specifications are in the README.md file and demos are available. In Figures \ref{fig:delta_ml_vs_massprecision} and \ref{fig:mcmc}, we demonstrate the code's capabilities with a hypothetical brown dwarf and background star of sufficient brightness to detect microlensing events. We find the astrometric microlensing shift of the background star over time and perform a Monte Carlo simulation to find the mass uncertainty. Figure \ref{fig:mcmc} shows that in favorable conditions, with a close approach by the brown dwarf, good constraints can be placed on the brown dwarf mass.

\section{Summary and Conclusions} \label{conc}

We developed a new method to search for predicted astrometric microlensing events. Our method searches for events that will lead to a target mass constraint on the lens, rather than a threshold astrometric signal. The main advantage of our method is that its does not require a lens mass estimate. Our method only requires a lens mass precision goal and a given astrometric precision. This allows us to search for events that provide useful mass constraints for a given object. Although we have applied this method to brown dwarf lenses in this study, our method can be generalized to all types of possible lenses. Along with this paper, we make the code implementing our new method publicly available to the community.   

We used our new method to search for brown dwarf microlensing events occurring between 2014-2032. We selected 1905 brown dwarfs in the Solar Neighborhood from \cite{kirkpatrick2021}, \cite{best}, and \cite{ultracool_sheet}, and background stars from the Legacy Survey \citep{dey}. We then used the position and motion uncertainties to perform a Monte Carlo simulation to find astrometric microlensing events that would allow a brown dwarf lens's mass to be determined with $\approx 10M_{\text{jup}}$ precision. No events meeting this criteria were found. We attribute this lack of events to a low expected rate of events per year, $\sim10^{-5}$, for nearby brown dwarfs in the Legacy Survey fields.

For future predicted microlensing events of brown dwarfs, we recommend searching the Galactic Plane to increase the number of brown dwarf lenses in a crowded field. Current nearby brown dwarf catalogs are incomplete by $\sim$ 15\% in this region \citep{kirkpatrick2021}, primarily due to source confusion in the 2MASS and WISE catalogs. Targeted searches for this missing population of brown dwarfs could find objects that are significantly more likely to cause an astrometric microlensing event. We found that although deeper imaging outside the Galactic Plane can increase the background source density, this increase is two orders of magnitude smaller than the increase in background source density in and around the Galactic Plane compared to fields outside the Galactic Plane. 

Astrometric microlensing provides the only currently known method to directly measure the mass of single, free-floating brown dwarfs, independent of atmospheric and evolutionary models. The upcoming Vera C. Rubin Observatory and Roman Space Telescope could provide the means to both find new nearby brown dwarf lenses and follow up predicted events. Our methodology applied in the Galactic Plane, combined with current or upcoming surveys means that direct measurement of brown dwarf masses via astrometric microlensing is likely in reach in the near future.  

\section*{Acknowledgments}
The authors thank J. Davy Kirkpatrick for useful discussions. E.C.M. is supported by the Heising Simons Foundation 51 Pegasi b Fellowship (\# 21-0684). 

The Legacy Surveys consist of three individual and complementary projects: the Dark Energy Camera Legacy Survey (DECaLS; Proposal ID \#2014B-0404; PIs: David Schlegel and Arjun Dey), the Beijing-Arizona Sky Survey (BASS; NOAO Prop. ID \#2015A-0801; PIs: Zhou Xu and Xiaohui Fan), and the Mayall z-band Legacy Survey (MzLS; Prop. ID \#2016A-0453; PI: Arjun Dey). DECaLS, BASS and MzLS together include data obtained, respectively, at the Blanco telescope, Cerro Tololo Inter-American Observatory, NSF’s NOIRLab; the Bok telescope, Steward Observatory, University of Arizona; and the Mayall telescope, Kitt Peak National Observatory, NOIRLab. The Legacy Surveys project is honored to be permitted to conduct astronomical research on Iolkam Du’ag (Kitt Peak), a mountain with particular significance to the Tohono O’odham Nation.

NOIRLab is operated by the Association of Universities for Research in Astronomy (AURA) under a cooperative agreement with the National Science Foundation.

This project used data obtained with the Dark Energy Camera (DECam), which was constructed by the Dark Energy Survey (DES) collaboration. Funding for the DES Projects has been provided by the U.S. Department of Energy, the U.S. National Science Foundation, the Ministry of Science and Education of Spain, the Science and Technology Facilities Council of the United Kingdom, the Higher Education Funding Council for England, the National Center for Supercomputing Applications at the University of Illinois at Urbana-Champaign, the Kavli Institute of Cosmological Physics at the University of Chicago, Center for Cosmology and Astro-Particle Physics at the Ohio State University, the Mitchell Institute for Fundamental Physics and Astronomy at Texas A\&M University, Financiadora de Estudos e Projetos, Fundacao Carlos Chagas Filho de Amparo, Financiadora de Estudos e Projetos, Fundacao Carlos Chagas Filho de Amparo a Pesquisa do Estado do Rio de Janeiro, Conselho Nacional de Desenvolvimento Cientifico e Tecnologico and the Ministerio da Ciencia, Tecnologia e Inovacao, the Deutsche Forschungsgemeinschaft and the Collaborating Institutions in the Dark Energy Survey. The Collaborating Institutions are Argonne National Laboratory, the University of California at Santa Cruz, the University of Cambridge, Centro de Investigaciones Energeticas, Medioambientales y Tecnologicas-Madrid, the University of Chicago, University College London, the DES-Brazil Consortium, the University of Edinburgh, the Eidgenossische Technische Hochschule (ETH) Zurich, Fermi National Accelerator Laboratory, the University of Illinois at Urbana-Champaign, the Institut de Ciencies de l’Espai (IEEC/CSIC), the Institut de Fisica d’Altes Energies, Lawrence Berkeley National Laboratory, the Ludwig Maximilians Universitat Munchen and the associated Excellence Cluster Universe, the University of Michigan, NSF’s NOIRLab, the University of Nottingham, the Ohio State University, the University of Pennsylvania, the University of Portsmouth, SLAC National Accelerator Laboratory, Stanford University, the University of Sussex, and Texas A\& M University.

BASS is a key project of the Telescope Access Program (TAP), which has been funded by the National Astronomical Observatories of China, the Chinese Academy of Sciences (the Strategic Priority Research Program “The Emergence of Cosmological Structures” Grant \# XDB09000000), and the Special Fund for Astronomy from the Ministry of Finance. The BASS is also supported by the External Cooperation Program of Chinese Academy of Sciences (Grant \# 114A11KYSB20160057), and Chinese National Natural Science Foundation (Grant \# 11433005).

The Legacy Survey team makes use of data products from the Near-Earth Object Wide-field Infrared Survey Explorer (NEOWISE), which is a project of the Jet Propulsion Laboratory/California Institute of Technology. NEOWISE is funded by the National Aeronautics and Space Administration.

The Legacy Surveys imaging of the DESI footprint is supported by the Director, Office of Science, Office of High Energy Physics of the U.S. Department of Energy under Contract No. DE-AC02-05CH1123, by the National Energy Research Scientific Computing Center, a DOE Office of Science User Facility under the same contract; and by the U.S. National Science Foundation, Division of Astronomical Sciences under Contract No. AST-0950945 to NOAO.

\bibliography{references}

\end{document}